\documentstyle[twocolumn,aps]{revtex}

\begin{document}

\hyphenation{Ka-pi-tul-nik}

\twocolumn[
\hsize\textwidth\columnwidth\hsize\csname@twocolumnfalse\endcsname
\draft
\title{True Superconductivity in a 2D "Superconducting-Insulating" System}
\author{Nadya Mason and Aharon Kapitulnik}
\address{Departments of Applied Physics and of Physics, Stanford
University, Stanford, CA 94305, USA}
\date{\today}
\maketitle

\begin{abstract}
We present results on disordered amorphous films which are expected
to undergo a field-tuned Superconductor-Insulator Transition.
Based on low-field data and I-V characteristics, we find evidence
of a low temperature Metal-to-Superconductor transition. This
transition is characterized by hysteretic magnetoresistance and 
discontinuities in the I-V curves. The metallic phase just above 
the transition is different from the "Fermi Metal" before superconductivity sets in.

\end{abstract}

\pacs{PACS numbers: 74.20.M, 74.76.-W, 73.40.Hm }
]

	The two-dimensional (2D) Superconductor-Insulator (SI) Transition
has attracted widespread attention in recent years
\cite{goldman1}. Initially, theory argued for the existence of a
magnetic field tuned, continuous phase transition from a
superconducting to an insulating state, with a quantum critical
point and corresponding quantum critical scaling about a critical
field $H_{SI}$ \cite{sondhi,fisher1,fisher2}. Experiments seemed
to confirm this scenario, although there was some concern that
the apparent critical exponents resembled those expected from
classical percolation \cite{aharony} and that the critical
resistance at the transition was not the quantum of resistance
for Cooper pairs, $h/4e^2$ \cite{yazdani,hebard,goldman}.
However, recent  experiments \cite{ephron,mason,valles} have
challenged the general existence of such a transition,
demonstrating that the apparent transition is only a crossover to
a new metallic state at low  temperatures. Similar results have
been obtained on quantum Hall liquid-to-insulator transitions and
on arrays of Josephson junctions; both are presumed to be in the
same universality class as the 2D SI transition
\cite{zant,shahar,liu}. The question of whether  a true
superconducting phase is possible within this new experimental
situation was an unresolved issue.

	In this paper, we  present results on magnetoresistance and
current-voltage measurements carried out on 2D  films of MoGe
below $H_{SI}$. Observations of a dramatic drop in resistance at a
critical magnetic  field $H_{SM} << H_{SI}$, along with
corresponding hysteresis near that field and instabilities in I-V
curves, point to the existence of a low-field quantum phase
transition, possibly a first order one, to a true superconducting
state. The metallic phase above the transition is not a normal
Fermi Liquid, at least in any simple sense, because its resistance
is orders of magnitudes smaller than the normal state ("Fermi")
resistance, it has very large magnetoresistance, and it exhibits
non-linear transport unlike a conventional metal. The
superconducting state can be characterized by vortices (far
separated due to the low field) that are pinned on impurities to
provide the true zero resistance.

	Samples for which we present data in this paper are 30$\dot{A}$ and
40$\dot{A}$ $Mo_{43}Ge_{57}$ thin films, sandwiched  between insulating
layers of amorphous Ge on SiN substrates. The 30$\dot{A}$
and 40$\dot{A}$ samples have sheet  resistances at 4.2K of $R_{N} \sim$1300
$\Omega / \Box$ and $R_{N} \sim$800 $\Omega / \Box$,
respectively; $T_C$'s of 0.5K and  1K; $H_{C2}$'s of 1.4T and 1.9T. Previous
studies have determined the films to be amorphous and
homogeneous on  all relevant length scales. We patterned the films into
4-probe structures, and measured them in a dilution
refrigerator using standard low-frequency lock-in techniques. Data was
taken at a measurement frequency of  ${\it f}_{AC}=27.5Hz$
with an applied bias of 1nA (well within the Ohmic regime). Current-voltage
characteristics were measured as dV/dI curves, using
battery-operated electronics to add a slow DC ramp  voltage to a
lockin AC output.

	Figure 1 shows a magnetoresistance curve for a 40$\dot{A}$  sample,
for temperatures 50mK and 200mK. The field and temperature
dependencies at higher fields were examined previously on  similar
samples \cite{yazdani,ephron,mason}. The temperature-independent "crossing
  point" -- apparent in both the main figure and in more detail in the
inset -- is expected from scaling  theories\cite{fisher1}, and is
of similar magnitude and quality as that obtained previously on
similar  samples. Below the crossing point, the temperature
curves spread and enter an "activated" regime, where $R \sim
e^{U(H)/T}$ and the derived activation energy, U(H), is
consistent with $U(H)=U_{0}Ln(H_{0}/H$),  a form expected in the
collective creep regime of vortices\cite{larkin} (here $U_0$ is of
order of  dislocation energy and $H_0$ is approximately
$H_{c2}$). At lower fields, the different temperature  curves
collapse onto each other, with the lower temperatures collapsing
at higher fields (this occurs at  750 Oe for 100mK, and 500 Oe
for 200mK, for example): this collapse marks where the system
enters a temperature-independent regime previously associated with
quantum tunneling\cite{ephron}. While it was previously unclear
whether this "metallic" region (of finite, temperature-independent
resistance) persisted to zero temperature, it is now evident that
the system enters a new phase at very low fields. Near 0.1 T,
the  resistance suddenly drops by more than 3 orders of
magnitude, approaching zero resistance to within the limits  of
our measurement. As seen from Figure 2, this drop is  best fit by
$R \sim 60(H-0.085)^\mu$ with  $\mu \sim 1$.  A kink in the
magnetoresistance interrupts the power law and a true zero
resistance  state seems to  occur below $\sim$ 185 Oe (see inset
of figure 2). Additional data taken for  other films of different
$R_{N}$ showed similar behavior. Previous experiments have
determined that  resistance saturation of lower $R_{N}$ films
seems to occur at temperatures below our measurement
capabilities\cite{ephron}. However, magnetoresistance curves for
these films do show the same qualitative  behavior as the higher
resistance films, with the data shifted to higher fields and
lower resistance values.

	To better examine the low field superconducting behavior, we  took more
sensitive resistance measurements for a small field range around zero.
Figure 3 shows  magnetoresistance measurements at 100mK for field up and
down sweeps from -600 to 600 Oe. The data clearly  shows an increase from
zero to finite resistance around 185 Oe -- evidence of a critical field,
$H_{SM}$,  for the transition to a true superconducting state. The up and
down sweeps are symmetric around  zero above $\sim$400 Oe, but asymmetric
and hysteretic below. The curves are independent of temperature below 100mK,
and are not affected by changes in bias current to within 2 orders of
magnitude. The value of the critical field corresponds to a vortex separation of
$\sim 5\xi _0 - 7\xi _0$, where  $\xi_0$ is the vortex core size.

	Further evidence of a low field phase transition to a
superconducting state is evinced by the dV/dI curves. Figure 4
shows typical dV/dI curves for a 30$\dot{A}$ film  at 20mK and
fields of 0.2 and 0.1 Tesla. The 0.2 T curve is at the end of the
"flattened"  region of the magnetoresistance curve, and has a
temperature dependence consistent with quantum tunneling of
vortices. The peak, evident at $\sim$1.2$\mu$A with a value
almost four times the normal state resistance, is typical for
both vortices in the flux-flow regime (see e.g.
\cite{ryu,hellerqvist})  and Josephson Junctions.  Examination of
the high current regime suggests that the system's behavior is
more similar to that of Josephson Junctions than to flux flow
vortices, since the leveling resistance  at high bias current is
approximately the normal state resistance (i.e., more than 10
times the calculated  flux flow resistance). At low fields and
high currents the sample seems to enter a new regime: the
structure evident in the 0.1 Tesla curve  -- peaks in dV/dI, or
discontinuities in I-V -- manifests sample behavior near $H_{SM}$.
This curve is both reproducible and hysteretic. Discontinuities
in I-V characteristics are likely due to vortex jumps and local
heatings. This can be caused by local inhomogeneities in the
sample, possibly phase separation into regions  with different
critical currents. This behavior has been seen in other systems,
also near quantum phase transitions\cite{valles}.

	The above results clearly point to a new physical situation
of the superconducting film at low temperatures and magnetic
fields below the upper critical field. For instance, we observed a
wide range of magnetic fields for which the system is a metal at
$T=0$. The metallic phase stabilizes at very low temperatures,
and is far from being a simple extrapolation of the normal state
"Fermi-metal" that we observed just above the bulk transition
temperature. This new metal is characterized by very low
resistance; as can be seen from Figure 1, at 0.2 T this resistance
is more than two orders of magnitude below the normal state
resistance. At that field the resistance is temperature
independent below $\sim 150 mK$ \cite{mason}. Furthermore, this
unusual metal has very strong magnetoresistance, especially very
close to the true superconducting transition $H_{SI}$. The
transport is different from a conventional metal in that the I-V
are non-linear at relatively low currents. The overall shape of
the I-V characteristics resembles that of a resistively shunted
Josephson-junction. The transition into the superconducting state
is also unusual because it is strongly hysteretic. Such behavior
is clear evidence for the existence of vortices. The origin of
the hysteresis could be due to either a genuine first order
transition to the true superconducting state, or else a dynamical
consequence of a glassy state in which the relatively low density
of vortices are frozen. In the latter case, this would be the
first observation of the long sought vortex-glass phase in 2D
superconductors \cite{fisher2}.

	Typical theoretical treatment of the SI system is to map it
onto the  so-called "dirty-boson" model, which considers bosons
interacting in the presence of disorder \cite{sondhi}. This model
predicts that for a field tuned transition with an arbitrary
amount of disorder a true  superconducting state exists at T=0,
when vortices are localized into a vortex-glass phase and Cooper
pairs are delocalized \cite{fisher1}. Above a critical field
$H_{C} = H_{SI}$, vortices are delocalized and Cooper  pairs
localize into an insulating Bose-glass phase. A Bose-metal, with
universal sheet resistance, should exist at the critical
resistance \cite{fisher2}. Most "dirty-boson" analyses predict a
continuous transition. A scaling analysis has been suggested
\cite{fisher1} which seems to yield a good fit to experimental
data in a limited range of temperatures and fields
\cite{yazdani,hebard,goldman}. However, results of the type
presented in this paper, as well as others
\cite{ephron,mason,valles}, cast doubt on the generality of the
theory. Other theories also predicted a  pure SI transition, even
in the presence of dissipation, and did not allow for a range of
a metallic phase \cite{chakravarty1,chakravarty2,wagenblast}. The
possibility of a metallic phase intervening between the
insulating and superconducting phases was first proposed by Mason
and Kapitulnik \cite{mason}, who suggested a new phase diagram
for the SI system in which a superconductor-metal transition
exists and depends on a new parameter, $\alpha$, which describes
a coupling to dissipation \cite{mason}. This idea can be
connected to a theory by Shimshoni {\it et al.} \cite{shimshoni}
which explains low temperature metallic states in SI systems as
an effect of dissipative quantum tunneling of vortices. In this
model, the SI transition is percolation-like, with couplings
between superconducting "puddles" in the superconducting phase and
insulating "puddles" in the insulating phase. A different
approach to obtaining a puddle-like structure was proposed by
Spivak and Zhou \cite{spivak}. In that paper it was argued that
mesoscopic fluctuations  of the order parameter at very low
temperatures manifest themselves in multiple re-entrant
transitions between superconducting and normal states; hence, the
creation of puddles for which global superconductivity is
obtained via an effective random SNS junctions array. Mason and
Kapitulnik \cite{mason} also showed that at higher temperatures
the system almost undergoes a superconductor-insulator transition
with a correlation length exponent very close to that of
classical percolation. This observation further  strengthened the
proposals \cite{shimshoni,spivak} that the system breaks into
puddles that almost connect via a classical percolation process
before settling into a new metallic phase that is  dominated by
vortex dissipation. The Josephson-Junction-like I-V
characteristics are perhaps the most  striking evidence that
indeed the system breaks down into domains which are connected
via Josephson tunneling. Further analysis of a puddle model
consisting of strongly fluctuating superconducting grains
embedded in a metallic matrix led Spivak {\it et al.}
\cite{spivak1} to predict a metal-to-superconductor transition
with a metallic phase just above the transition which is
dominated by Andreev reflections from the almost superconducting
grains. The resistance of such a phase has to be much lower than
the "normal"  resistance of the system, an occurrence that has
consistently been observed in our samples. Another approach to a
metallic phase in an SI system was recently taken by Dalidovich
and Phillips \cite{phillips2} who showed that dissipation causes
the metallic phase and ultimately is responsible for the
insulating phase; in this case, a true superconducting state is
expected as T vanishes.

In summary, we presented in this paper evidence of a genuine
transition between a new metallic state and a superconducting
state in 2D films. We believe that our experiment presents the
first evidence for a T=0 quantum phase transition of this kind.
While a simple phenomenology based on a "puddle" model of
superconducting and metallic regions can qualitatively explain
the main features of our experiment, more work is needed to fully
understand the nature of the new metallic state and the
superconducting transition.\\

We thank David Ephron whose thesis work motivated parts of this study.
We thank Steve Kivelson and Boris Spivak for many useful discussions.
We especially thank Steve Kivelson for critical reading of the manuscript.  
Work supported by NSF/DMR. NM thanks Lucent CRFP Fellowship program for 
support.  Samples prepared at Stanford's Center for Materials Research.

\figure{FIG. 1. Magnetoresistance of a 40$\dot{A}$ sample at 200 and 50 mK.
Inset (marked on the main figure as a dashed box) shows the high field
portion with the crossing point.
\label{fig1}}

\figure{FIG. 2. Low field portion of the magnetoresistance shown in
Fig.1. Dashed line represents a linear fit with an intersection field
of 850 Oe. The inset shows the actual critical field of the sample of 185 Oe.
\label{fig2}}

\figure{FIG. 3. Magnetoresistance near the critical field of the
sample. Arrows show the direction of the field sweep.  Curves are all
shifted by 87 Oe to account for trapped flux in the 16T magnet.
\label{fig3}}

\figure{FIG. 4. Dynamic resistance of a 30$\dot{A}$ sample at 200 Oe
and 100 Oe.
\label{fig4}}

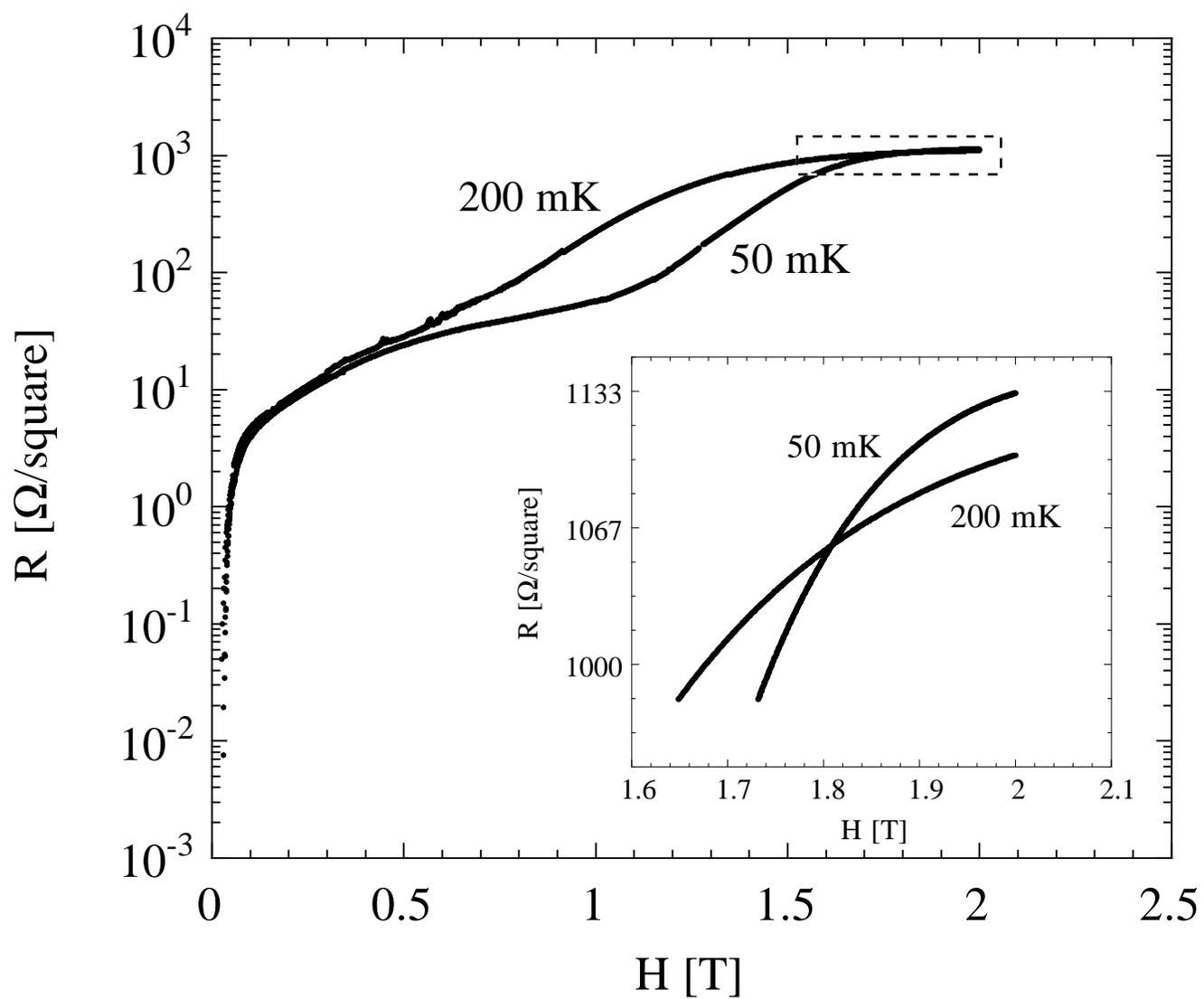
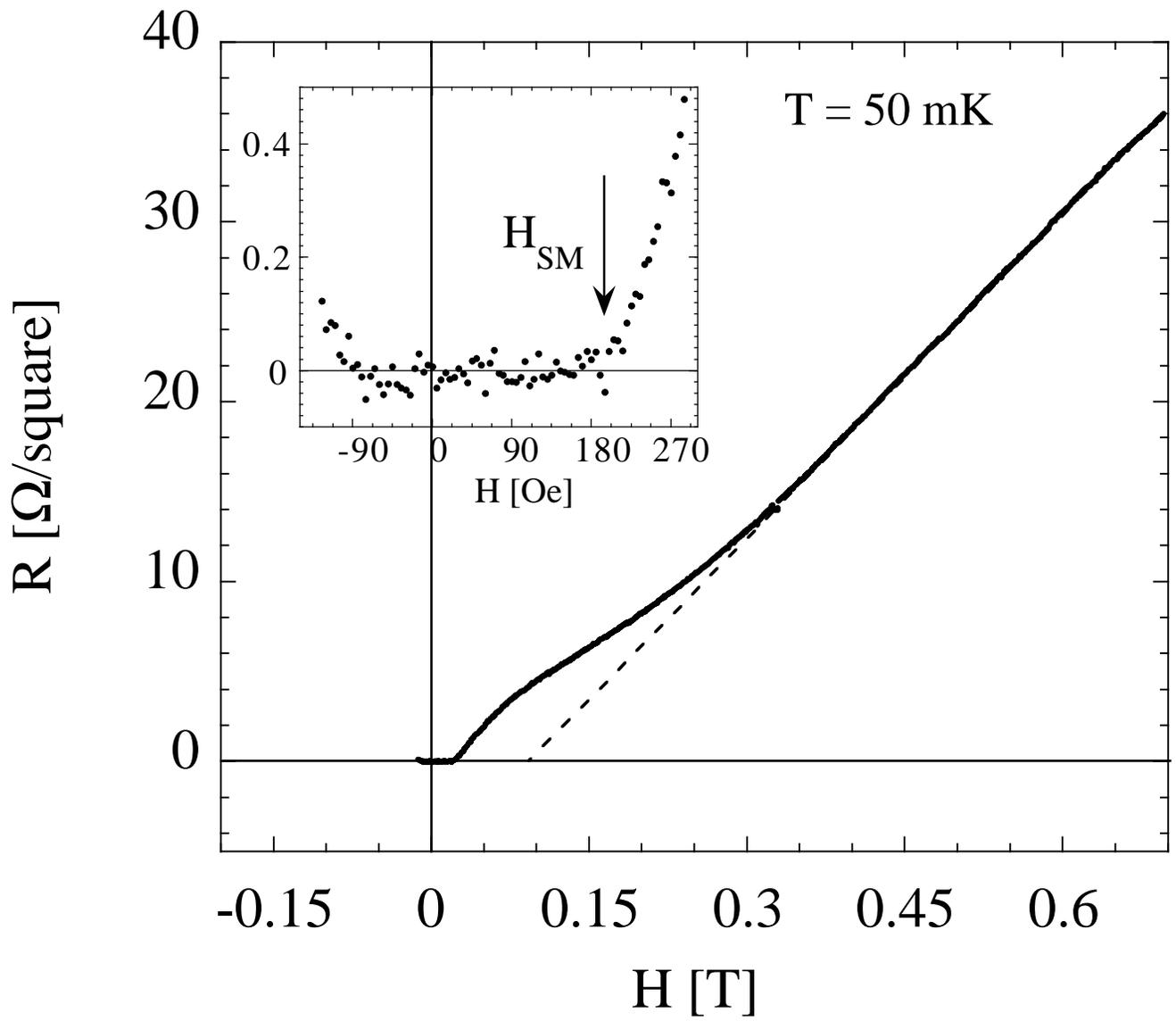
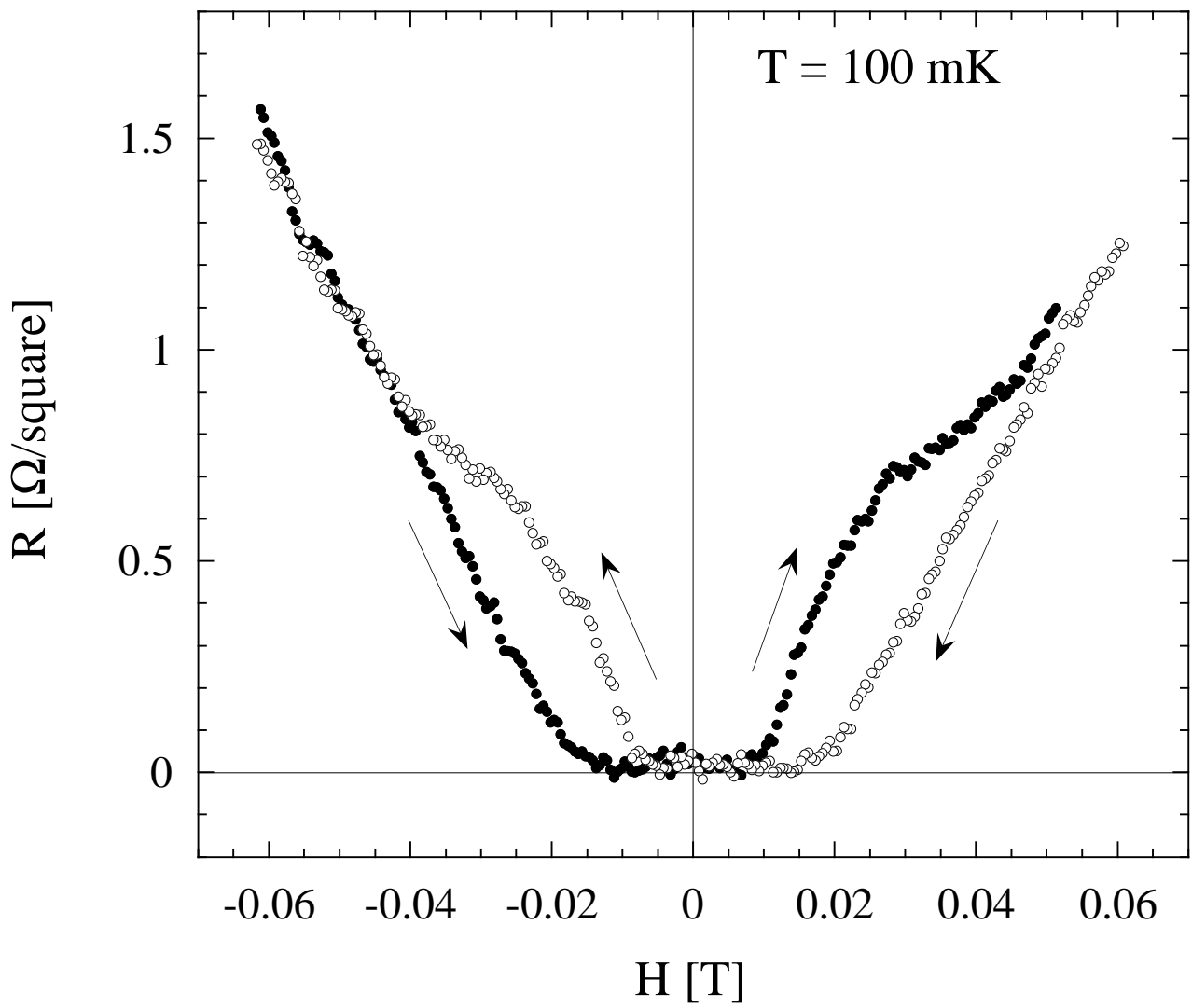
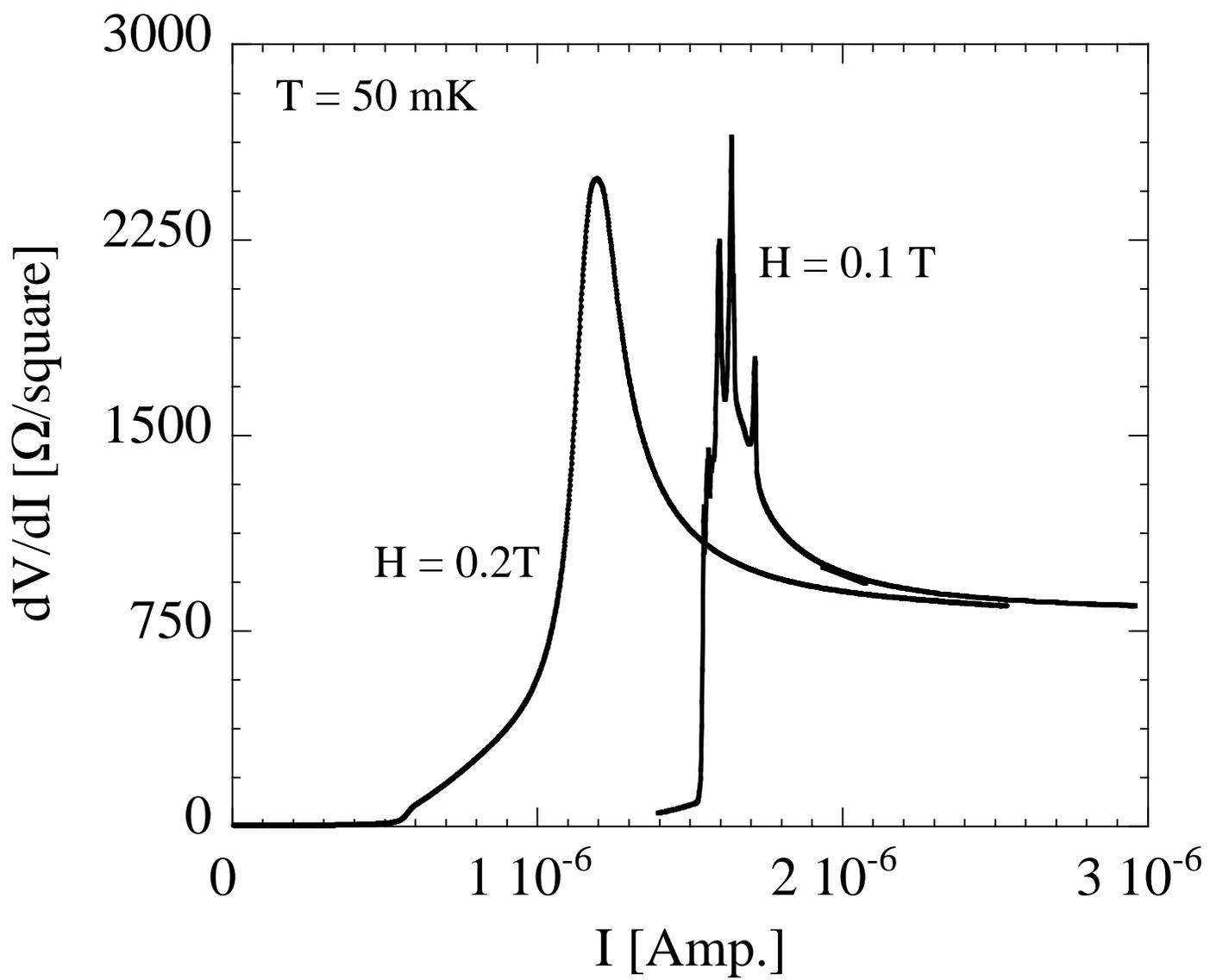
\end{document}